\RequirePackage{fix-cm}
\documentclass[twocolumn]{svjour3}          
\smartqed  
\usepackage{graphicx}
\usepackage[backend=biber, style=phys, sorting=none, doi=true, eprint=true, biblabel=brackets,maxbibnames=6, minbibnames=6]{biblatex}
\usepackage{csquotes}

\AtEveryBibitem{%
  \clearfield{Extra}%
}

\AtEveryBibitem{%
  \clearfield{note}%
}

\AtEveryBibitem{%
  \clearfield{series}%
}

\AtEveryBibitem{%
  \clearfield{doi}%
}

\AtEveryBibitem{%
  \clearlist{language}%
}

\usepackage{textalpha}
\DeclareUnicodeCharacter{2212}{-}
\DeclareUnicodeCharacter{2009}{ }
\DeclareUnicodeCharacter{2248}{$\approx$}
\DeclareUnicodeCharacter{202F}{ }
\DeclareUnicodeCharacter{223C}{ }
\DeclareUnicodeCharacter{2086}{$_6$}
\DeclareUnicodeCharacter{2084}{$_4$}
\DeclareUnicodeCharacter{2082}{$_2$}
\DeclareUnicodeCharacter{2083}{$_3$}

\usepackage{amsmath,amssymb}
\usepackage{cuted}
\usepackage{flushend}
\AfterEndEnvironment{strip}{\leavevmode}
\usepackage{silence}
	\ErrorsOff[latex]

\DefineBibliographyStrings{english}{
	andothers = {\mkbibemph{et al.}}
}

\usepackage{amsmath}
\renewcommand{\thesection}{\Roman{section}.} 
\renewcommand{\thesubsection}{\Alph{subsection}.}

\usepackage{titlesec}
\usepackage{etoolbox} 

\newbool{isFirstSection}
\setbool{isFirstSection}{true}

\pretocmd{\section}{%
  \ifbool{isFirstSection}{%
    \titlespacing*{\section}{0pt}{0pt}{0pt}%
    \setbool{isFirstSection}{false}%
  }{\titlespacing{\section}{0pt}{12pt}{0pt}
}%
}{}

\usepackage[colorlinks=true, pdfstartview=FitV, linkcolor=blue, citecolor=blue, urlcolor=blue,breaklinks=true]{hyperref}

\usepackage{xcolor,color,soul}

\AtBeginBibliography{\small}

\usepackage{placeins}

\begin{document}

\begin{strip}
\begin{center}
\vspace{-5ex}
{\Large \textbf{Evidence for superconductivity at 190 K\\ in a pressure-overdoped cuprate} \vspace{5pt}}
\\

{Alexander C. Mark$\mathrm{^a}$, Huu T. Do$\mathrm{^b}$, David Rodriguez$\mathrm{^a}$, D. Gonzalez Arevalo$\mathrm{^a}$, Adam Denchfield$\mathrm{^c}$, \\ Eduardo H. T. Poldi$\mathrm{^d}$, Daniel P. Phelan$\mathrm{^d}$, 
N. K. Man$\mathrm{^e}$, and Russell J. Hemley$\mathrm{^{d,f}}$ \vspace{5pt}} 
\\

$\mathrm{^a}$ \textit{Department of Physics, University of Illinois Chicago, Chicago IL 60607}\\
$\mathrm{^b}$ \textit{Institute of Theoretical and Applied Research, Duy Tan University, Hanoi 100000, Vietnam}\\
$\mathrm{^c}$ \textit{Center for Quantum Materials and Engineering, University of Texas at Austin, Austin TX 78712}\\
$\mathrm{^d}$ \textit{Materials Science Division, Argonne National Laboratory, Lemont, IL 60439}\\
$\mathrm{^e}$ \textit{School of Materials Science and Engineering, Hanoi University of Science and Technology, \\ Hanoi 100000,  Vietnam}\\
$\mathrm{^f}$ \textit{Departments of Physics, Chemistry, and Earth and Environmental Sciences, \\ University of Illinois Chicago, Chicago IL 60607}

\end{center}

\begin{abstract}
It is well established that the critical temperature ($T_c$) of cuprate superconductors can be tuned by pressure. For example, compression decouples the hole doping from chemical doping allowing for overdoped samples far beyond what is possible at ambient pressure. In this work, multiple techniques are used to probe the onset of the Meissner effect at $T_c$ as a function of pressure in $\mathrm{Pb_{0.4}Bi_{1.6}Sr_2Ca_2Cu_3O_{10+\delta}}$ (Bi-2223) to 60 GPa in different compression environments. Samples compressed under quasihydrostatic conditions exhibit a distinctive non-monotonic pressure dependence of $T_c$ below 25 GPa, in agreement with previous reports. With further increase in pressure $T_c$ climbs continuously to 190 K at 60 GPa. Evidence for critical temperatures exceeding those reported to date for cuprates at ambient and high pressures, the results may be understood in terms of the proposed second superconducting regime at high hole doping.
\vspace{1ex}
\hrule
\end{abstract}
\end{strip}

\section{Introduction}
\label{intro}
\vspace{\baselineskip}
The application of pressure has been a powerful tool for exploring high-temperature superconductivity in the \\ cuprates since their discovery \cite{bednorz_possible_1986}. One of the earliest observations was the large increase in the superconducting critical temperature ($T_c$) under moderate pressure in the La-Ba-Cu-O system, raising the  $T_c$ from 38 K to the then-unprecedented 52.5 K at 1.7 GPa \cite{chu_superconductivity_1987, chu_evidence_1987}. Pressure has continued to serve as a valuable tuning parameter to investigate the novel electronic properties in the cuprates (see Ref. \cite{mark_progress_2022} and references therein). Notably, several cuprates exhibit an unusual non-monotonic pressure dependence of $T_c$, with systematically higher critical temperatures than those observed at ambient pressure \cite{chen_enhancement_2010, mito_uniaxial_2017, deng_higher_2019}. $\mathrm{Bi_{2}Sr_2Ca_2Cu_3O_{10+\delta}}$ (Bi-2223) is an ideal system to probe the effects of high pressure on cuprate superconductivity due to its structural stability up to megabar ($<$100 GPa) pressures \cite{mark_structure_2023} and increasing $T_c$ above 25 GPa up to 40 GPa \cite{chen_enhancement_2010}. 

Despite this enhanced-superconductivity of Bi-2223 under pressure, discrepancies are apparent among previously reported $T_c$  data \cite{klotz_hydrostatic_1993, adachi_electrical_2001, chen_enhancement_2010, zhou_quantum_2022, ohkuma_probing_2026}, a result that correlates with the different compression environments of these experiments. The $T_c$ of samples compressed in non-hydrostatic media is reduced relative to those pressurized in sufficiently hydrostatic media, with the pressure dependence of $T_c$ forming a dome of roughly constant curvature with sample dependent maxima between 5 and 20 GPa \cite{klotz_hydrostatic_1993, zhou_quantum_2022}. The latter Bi-2223 samples exhibit an unusual pressure dependent $T_c$ with a minimum near 25 GPa and increasing $T_c$ upon further compression \cite{chen_enhancement_2010}. These observations are consistent with direct measurements that the structures of these cuprates are extremely susceptible to deviatoric stresses \cite{mark_structure_2023} inducing large strain gradients across samples that suppress and even destroy superconductivity \cite{ohkuma_probing_2026}.

Here, we present measurements that extend the pressure dependence of $T_c$ in Bi-2223 under quasihydrostatic compression up to 60 GPa. We utilize three complementary techniques to measure the onset of the Meissner effect in a diamond anvil cell (DAC): AC susceptibility \cite{goldfarb_alternating-field_1991}, a double modulated susceptibility technique \cite{timofeev_growth_1998, timofeev_improved_2002}, and radio frequency (RF) susceptibility \cite{semenok_transmission_2026} methods. Our results suggest the existence of a second superconducting regime at high pressure in Bi-2223 similar to that proposed for highly overdoped cuprates at ambient pressure \cite{geballe_enhanced_2009}. These experiments provide evidence for the highest critical temperatures in cuprates reported to date. 

\section{Methods}
\label{methods}
\vspace{\baselineskip}
\subsection{Sample Preparation}

\begin{figure}
\includegraphics[width=0.5\textwidth]{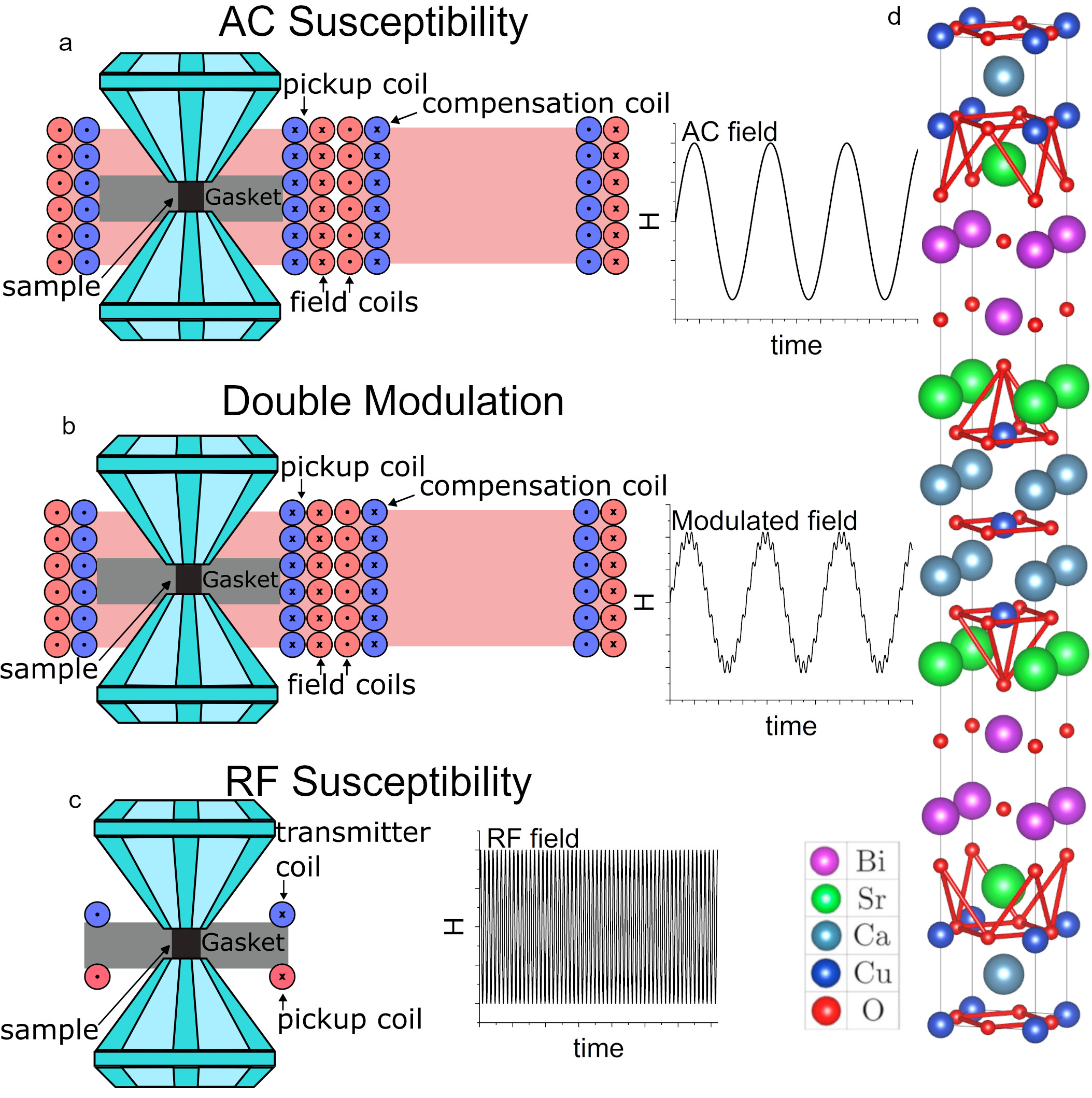}
\caption{DAC techniques used to measure signatures of superconductivity in Bi-2223 in this study. a) Setup for AC susceptibility measurements including a pickup and field coil around a diamond anvil. A matched reverse-wound compensation coil is placed next to the sensor to compensate for stray fields. $\cdot$ and x indicate coil is wound into or out of the page, respectively. Representative low frequency AC magnetic field waveform presented next to setup. b) Double modulation setup identical to the AC susceptibility setup, but utilizing a modulated AC field composed of a high amplitude low frequency modulation wave and a high frequency low amplitude excitation wave. c) RF susceptibility setup with single turn RF transmitter coil and pickup coil around DAC and high-frequency RF field. d) Unit cell of Bi-2223.}
\label{Fig:schematics} 
\end{figure}

Samples were synthesized using conventional solid state growth methods \cite{grivel_effects_1993, macmanus-driscoll_study_1997}. Precursor $\mathrm{Bi_2O_3}$, $\mathrm{PbO}$, $\mathrm{CuO}$, $\mathrm{SrCO_3}$, and $\mathrm{CaCO_3}$ were ground into fine powders and mixed together in the stoichiometric ratio corresponding to $\mathrm{Bi_{1.6}Pb_{0.4}Sr_2Ca_2Cu_3O_{10+\delta}}$ ($\delta = 0.16$). Pb was included in the compound to promote the formation of the Bi-2223 phase with negligible impact on $T_c$ \cite{takano_high-tc_1988, dou_effect_1995}. Samples were calcined at 800$^{\circ}$C and intermittently reground over 24 hours to form mainly $\mathrm{Bi_2Sr_2CaCu_2O_{8+\delta}}$ along with minority phases $\mathrm{Ca_2PbO_4}$, $\mathrm{Bi_2Sr_2Cu_2O_{6+\delta}}$ and $\mathrm{(Ca,Sr)_2CuO_3}$. Samples were then compressed into pellets at 5 tons/cm$^{2}$ and sintered at 855 $^{\circ}$C for 8 days. The resultant samples were found to be of comparable quality to those used in previous studies, with a near optimal $T_c$ = 109 K \cite{man_improvement_2019, man_signature_2024, gonzalez_arevalo_enhanced_2026}. $\delta$ was estimated to be 0.16 using the $T_c$-$\delta$ relation reported in Ref. \cite{presland_general_1991}.

Samples were loaded into multiple non-magnetic CuBe diamond anvil cells (DACs) with diamond culet sizes ranging from 100 to 300 $\mu$m and non-magnetic gaskets using various pressure transmitting media to give different degrees of quasihydrostaticity. Below 10 GPa a methanol / ethanol mixture was used as a fluid pressure transmitting medium.  For higher pressure experiments, powder mixtures of Bi-2223 and KCl in 20/80 and 50/50 mass ratios were used. Pressures were measured using the standard ruby fluorescence technique \cite{dewaele_compression_2008}. Above 30 GPa pressures were confirmed with diamond edge Raman measurements \cite{akahama_pressure_2006}. 

\begin{figure*}
\centering
\includegraphics[width=\textwidth]{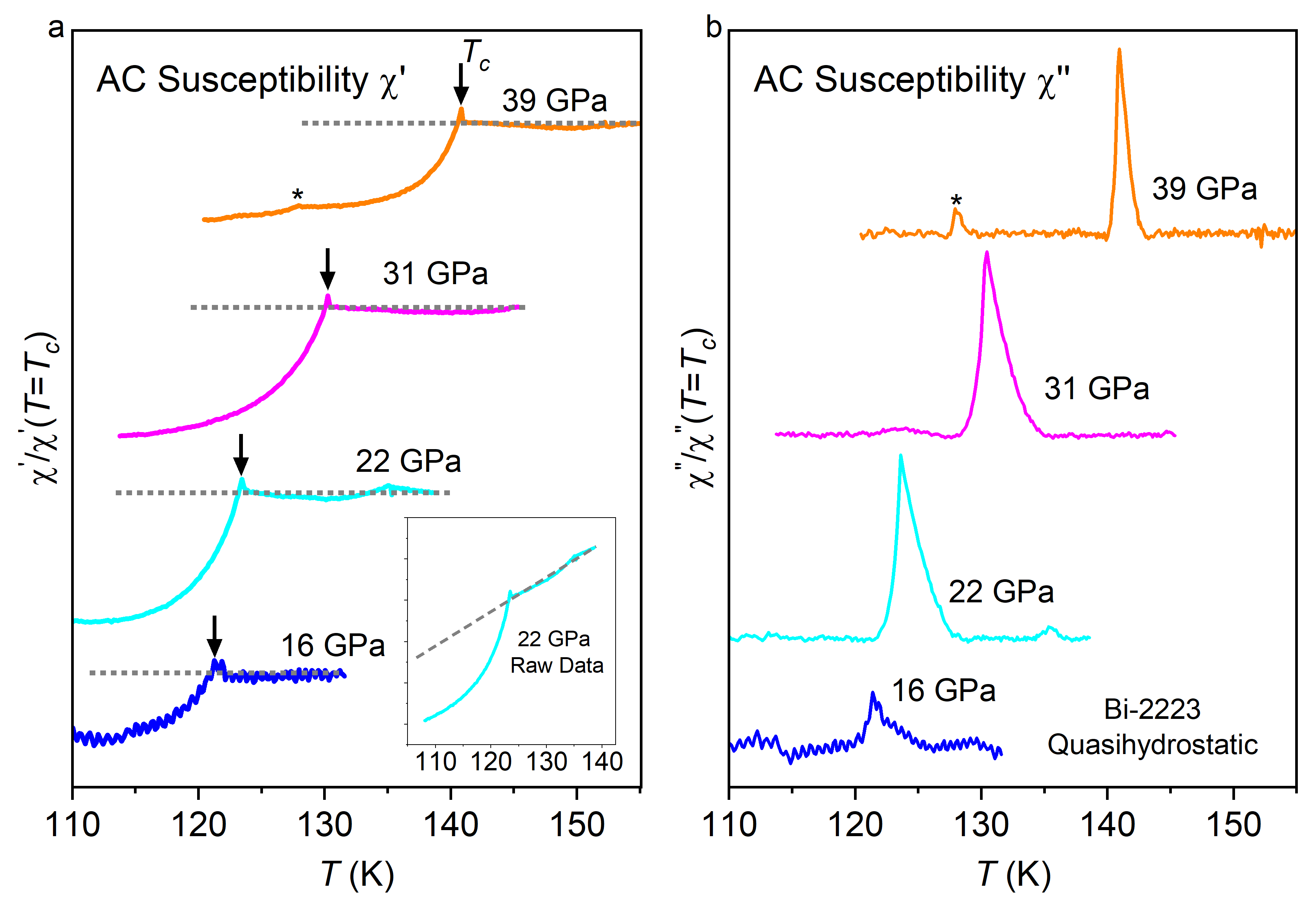}
\caption{a) $\chi'(T)$ for larger quasihydrostatic 20/80 Bi-2223/KCl samples. Inset: representative linear baseline subtraction near $T_c$. b) Corresponding $\chi''(T)$. Data collected at a driving frequency of $f$ = 98 Hz. Additional oscillations apparent in 16 GPa data due to low integration time used in demodulation. The background above $T_c$ was approximately linear for all runs, and a linear background has been subtracted around $T_c$ from all measurements to account for the diamagnetism of the pressure cells. Signals at 39 GPa (*) are attributed to pressure gradients developing across the sample, resulting in different regions having different $T_c$ values \cite{deemyad_dependence_2003}. Dotted lines are guides to the eye; signal traces are offset for clarity.}
\label{Fig:ACsusc} 
\end{figure*}

\begin{figure}
\centering
\includegraphics[width=0.45\textwidth]{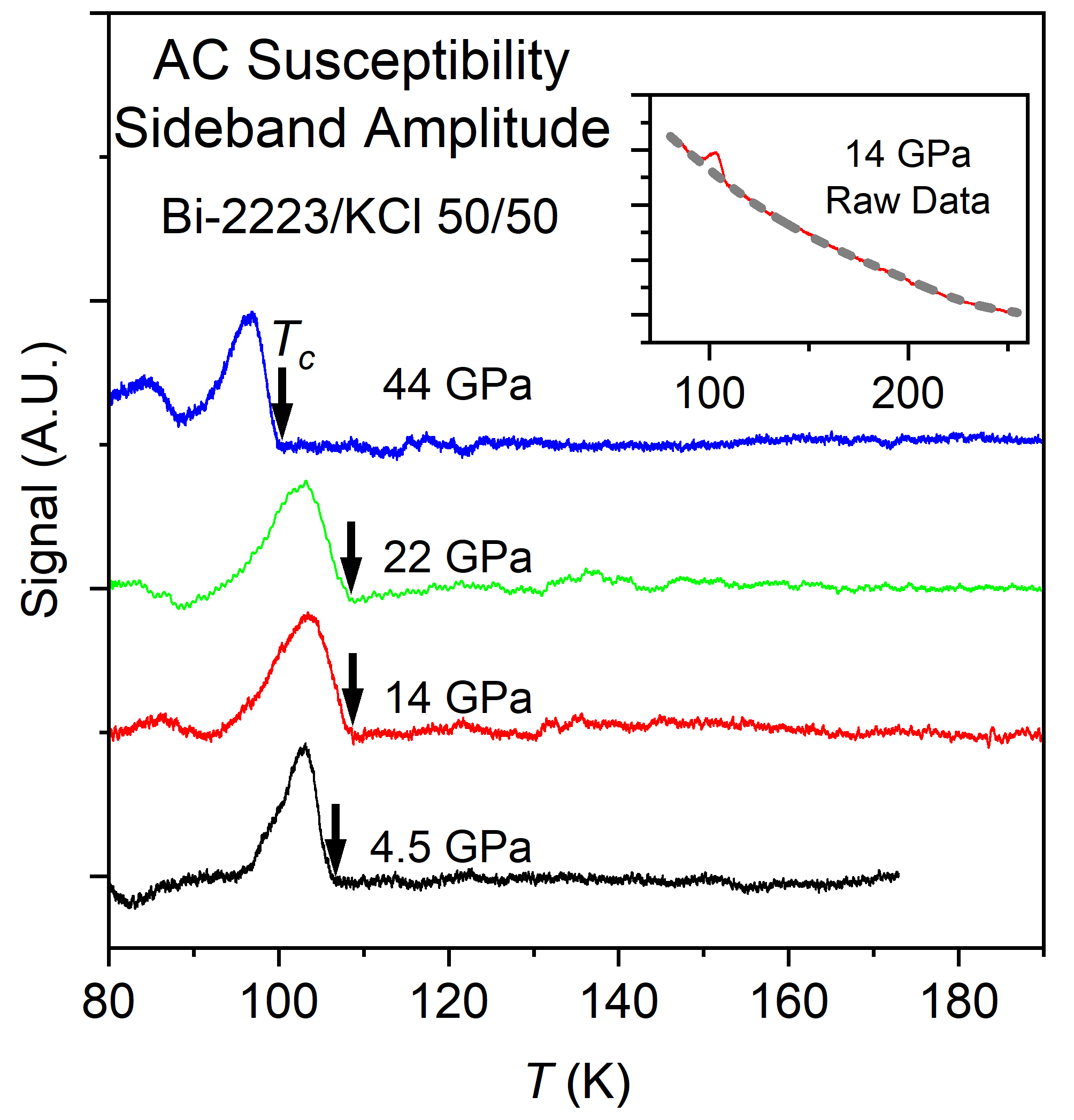}
\caption{Amplitudes of modulated signals from the less-quasihydrostatic 50/50 mixture of Bi-2223/KCl showing deviations from the linear background at $T_c$. Inset: representative background baseline (grey dashed line). Data offset for clarity.}
\label{Fig:NHS} 
\end{figure}

\begin{figure*}
\centering
\includegraphics[width=\textwidth]{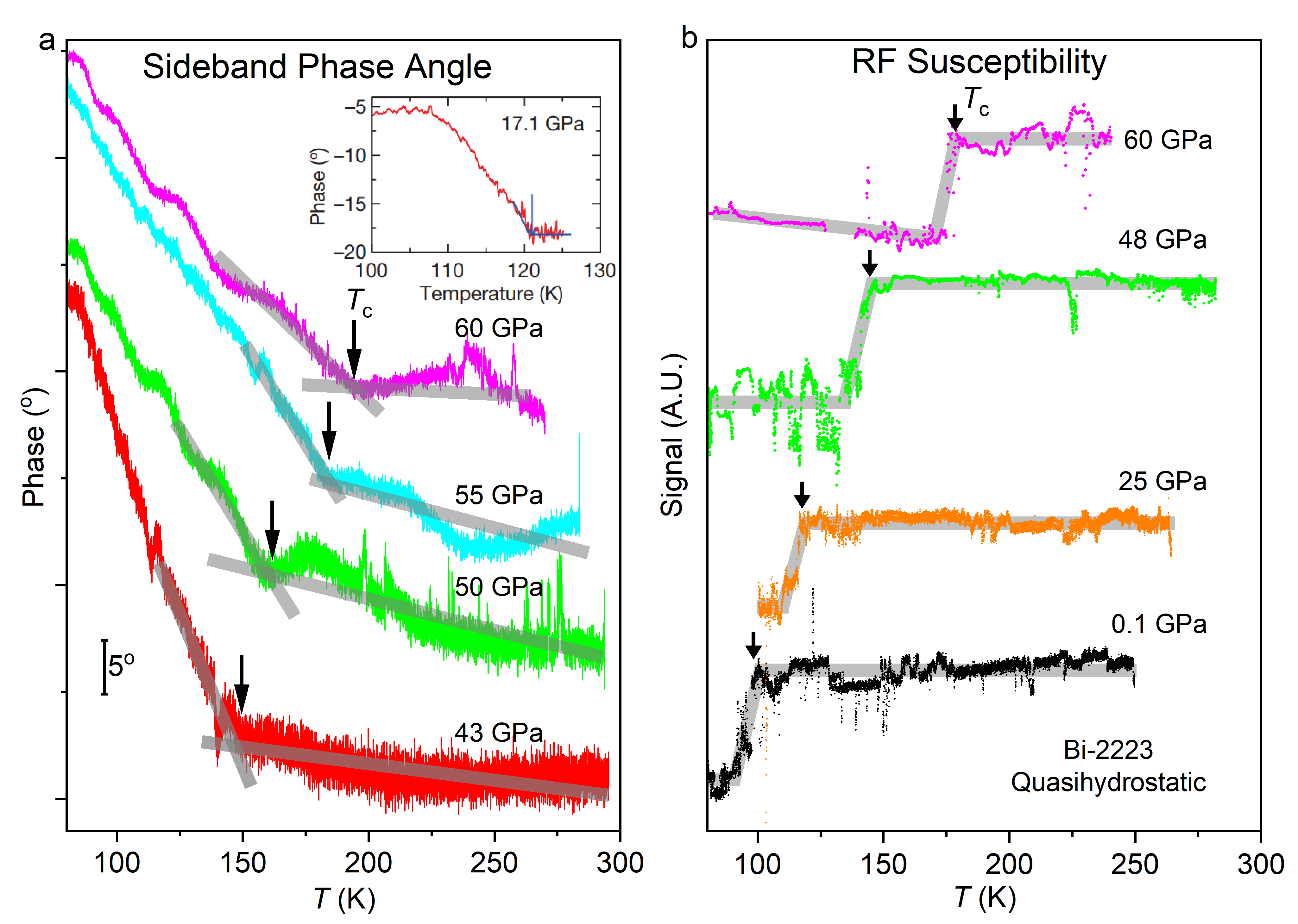}
\caption{a) Temperature dependence of the sideband phase angle of the quasihydrostatic 20/80 Bi-2223/KCl samples at various pressures. Data were collected using a driving frequency of 98 Hz and a modulation frequency of 2450 Hz. Arrows indicate assigned $T_c$ values using the criteria of Chen et al. \cite{chen_enhancement_2010}. Oscillations in measurements are instrumental and due to overtones of the driving frequencies. Data offset for clarity. Inset: phase angle - temperature plot adapted from Chen et al. \cite{chen_enhancement_2010} indicating the $T_c$ assignment criteria. b) Transmitted RF signals through a 20/80 Bi-2223/KCl mixture. Drops in the transmitted signal indicate diamagnetic transitions attributed to the onset of the Meissner effect. Linear baselines subtracted above $T_c$ in all runs and instrumental noise apparent in the 6 GPa dataset removed at 125 K - 150 K for clarity.}
\label{Fig:sideband-theta} 
\end{figure*}

\subsection{Electromagnetic Measurements}

Three different techniques were used to detect superconductivity in Bi-2223 by the onset of the Meissner effect as functions of pressure and temperature (Fig. \ref{Fig:schematics}). The circuitry for the AC susceptibility (Fig. \ref{Fig:schematics}a) and double modulation (Fig. \ref{Fig:schematics}b) methods were similar to those described in Ref \cite{timofeev_improved_2002}. For these methods, pickup and field coils were constructed from 48 AWG insulated Cu wire with an inner diameter of 12 mm. Individual coils of 420 turns with an average resistance of 100 $\mathrm{\Omega}$ and inductance of 40 $\mathrm{\mu}$H. Pickup coils were wound in matched pairs and connected in series with opposite polarity. Field coils were wound around the pickup coils in matched pairs, and connected in series with the same polarity. All coils were potted with a thin layer of Stycast 1622 clear epoxy and mounted on fiberglass supports. 

To detect changes in sample magnetic susceptibility (Fig. \ref{Fig:schematics}a) a standard lock-in based technique was used. Sinusoidal currents were sent through the outer excitation coils to generate a magnetic field. An instrumentation preamplifier buffered and amplified the signal induced in the inner pickup coils, and the signal was then demodulated with a SRS 860 lock-in amplifier. The components of the complex AC susceptibility $(\boldsymbol{\chi} = \chi' - i \chi'')$ are determined by $X(T) \propto  \chi'(T) \mathrm{cos}(\phi)$ and $Y(T) \propto -\chi''(T) \mathrm{sin}(\phi)$ where $\phi$ is a phase offset introduced by the measurement circuitry. 

For the double modulation technique, which was previously used to produce the first measurements of superconductivity at megabar pressures \cite{struzhkin_superconducting_1997-1, struzhkin_superconductivity_1997, timofeev_inductive_1999, timofeev_improved_2002}, current regulated signals were sent through the outer coils (Fig. \ref{Fig:schematics}b) to generate an AC magnetic field. The current signals have the form 

\begin{equation}
I(T)=|I_{ex}| \sin (2 \pi f_{ex} t) + |I_{m}| \sin (2 \pi f_{m} t),
\label{eq:input-current}
\end{equation}

\noindent where $I_{ex}$ and $I_{m}$ are the currents, and $f{ex}$ and $f_{mod}$ are the frequencies of the excitation and modulation signals, respectively. Corresponding signals from the pickup coils were then amplified by an instrumentation preamplifier and demodulated by the SRS 860 lock-in amplifier recording the lower sideband signal at $f_{SB} = f_{m}-2f_{ex}$. Typically, $|I_{ex}| < |I_{m}|$, resulting in a signal in the lock-in amplitude $\big[|V_{m}(T)|\big]$ and phase $\big[\theta_{m}(T)\big]$ at $f_{SB}$ when superconductivity is periodically quenched near $T_c$ due to the large modulating field. \cite{gregoryanz_superconductivity_2002}. For the smallest samples, driving signals were adjusted so that $|I_{ex}| = |I_{m}|$ to increase the signal-to-noise ratio (SNR) in $\theta_{m}(T)$, which deviates from the near constant background above $T_c$ as the sample begins to superconduct \cite{gregoryanz_superconductivity_2002, eremets_exploring_2003-1}. Instrumental artifacts are apparent in the recovered signals due to nonlinearities in the circuitry, induced background magnetic response, and the appearance of overtones and oscillations at multiples of the driving frequency. The critical temperature was assigned using the criteria of Chen et al. \cite{chen_enhancement_2010}, with $T_c$ identified as the temperature where the lock-in amplitude, $|V_{mod}(T)|$, exhibits the onset of a broad maximum and the phase angle, $\theta _{mod}(T)$, deviates from the near constant background. The technique was calibrated using ambient-pressure Bi-2223 samples with established $T_c$'s. 

Finally, an RF technique similar to that reported in Ref. \cite{semenok_transmission_2026} was used to measure the diamagnetic transition associated with superconductivity (Fig. \ref{Fig:schematics}c). These experiments utilized custom single-turn transformers on printed circuit boards with non-magnetic gaskets pressed into the center. RF signals were generated and demodulated using a SRS 844 lock-in amplifier at a driving frequency of 2.3 MHz.

\subsection{Density Functional Calculations}
Density functional theory (DFT) calculations with the PBE functional \cite{perdew_generalized_1996} using Quantum Espresso \cite{giannozzi_quantum_2009, giannozzi_advanced_2017} were used to estimate the change in $n_H$ with pressure for experimentally determined structures of optimally doped ($n_H$ = 0.16) Bi-2223 at 0, 15, 30, and 51 GPa \cite{mark_structure_2023}. Ultrasoft pseudopotentials as specified in the SSSP-precision database \cite{prandini_precision_2018} were used, with energy and density cutoffs of 320 and 80 Ry, respectively. A k-point mesh density of 0.15 $\mathrm{\AA}^{-1}$ was used for all calculations. The charge transfer was evaluated using the integral of the orbital PDOS, summing the contributions from all Cu and O orbitals within the $\mathrm{CuO_2}$ planes. We emphasize that while DFT calculations are historically a poor description of the fine-grained electronic structure of cuprates, we use the calculations to identify a qualitative trend regarding the pressure-driven difference of total charge density in the $\mathrm{CuO_2}$ planes.

\section{Results}
\label{results}

Figure \ref{Fig:ACsusc} shows the results of measurements conducted on quasihydrostatic compression using lock-in techniques to probe $\chi'$ and $\chi''$. $\chi'(T)$ signals show field expulsion at $T_c$ due to the Meissner effect, in agreement with  previous experiments at lower pressures in Ne \cite{chen_enhancement_2010}. At 39 GPa, the highest pressure where we directly measured the AC susceptibility with these techniques \cite{kim_system_1994}, multiple features in both $\chi '$ and $\chi ''$ are apparent. We attribute these features to the development of stress gradients across the sample as the compression environment stiffens, resulting in multiple superconducting transitions at different points across the sample. In contrast, comparable samples prepared in a 50/50 ratio of Bi-2223/KCl exhibited stark differences in their pressure dependence of $T_c$, with no increase in $T_c$ above 30 GPa (Fig. \ref{Fig:NHS}).

DACs loaded with 20/80 Bi-2223/KCl mass ratio exhibit $T_c$ pressure dependencies that agree with samples we compressed in methanol/ethanol, and with previous experiments using Ne as a medium at lower pressure \cite{chen_enhancement_2010}. For higher pressures using small samples, we utilized the double modulation technique \cite{timofeev_inductive_1999, timofeev_improved_2002} that has been previously applied up to 40 GPa for Bi-2223 \cite{chen_enhancement_2010}. Measurements of the temperature dependence of the phase angle were consistent with those of the side band amplitude, consistent with measurements on other systems \cite{shimizu_pressure-induced_1994, gregoryanz_superconductivity_2002}. We observe an abrupt change in $\mathrm{\Theta}(T)$ as a function of pressure that we identify as $T_c$, in agreement with previous reports \cite{chen_enhancement_2010}. As pressure is increased beyond the previous limit \cite{chen_enhancement_2010}, the data provides evidence that $T_c$ increases to 190 K at 60 GPa.

We used the RF technique to corroborate these measurements and to examine smaller samples at still higher pressures \cite{semenok_transmission_2026}. This method measures the reflection of RF signals from the superconducting material below $T_c$. Measurements performed on 20/80 Bi-2223 / KCl mixtures exhibit magnetic field expulsion in agreement with other techniques (Fig. \ref{Fig:sideband-theta} b), but $T_c$ values from RF methods were systematically 2-5 K lower than those observed using both the standard AC susceptibility setup and the double modulated technique, a difference that we attribute to backgrounds under the latter methods. 

\section{Discussion}
\vspace{\baselineskip}

\begin{figure}
\centering
\includegraphics[width=0.5\textwidth]{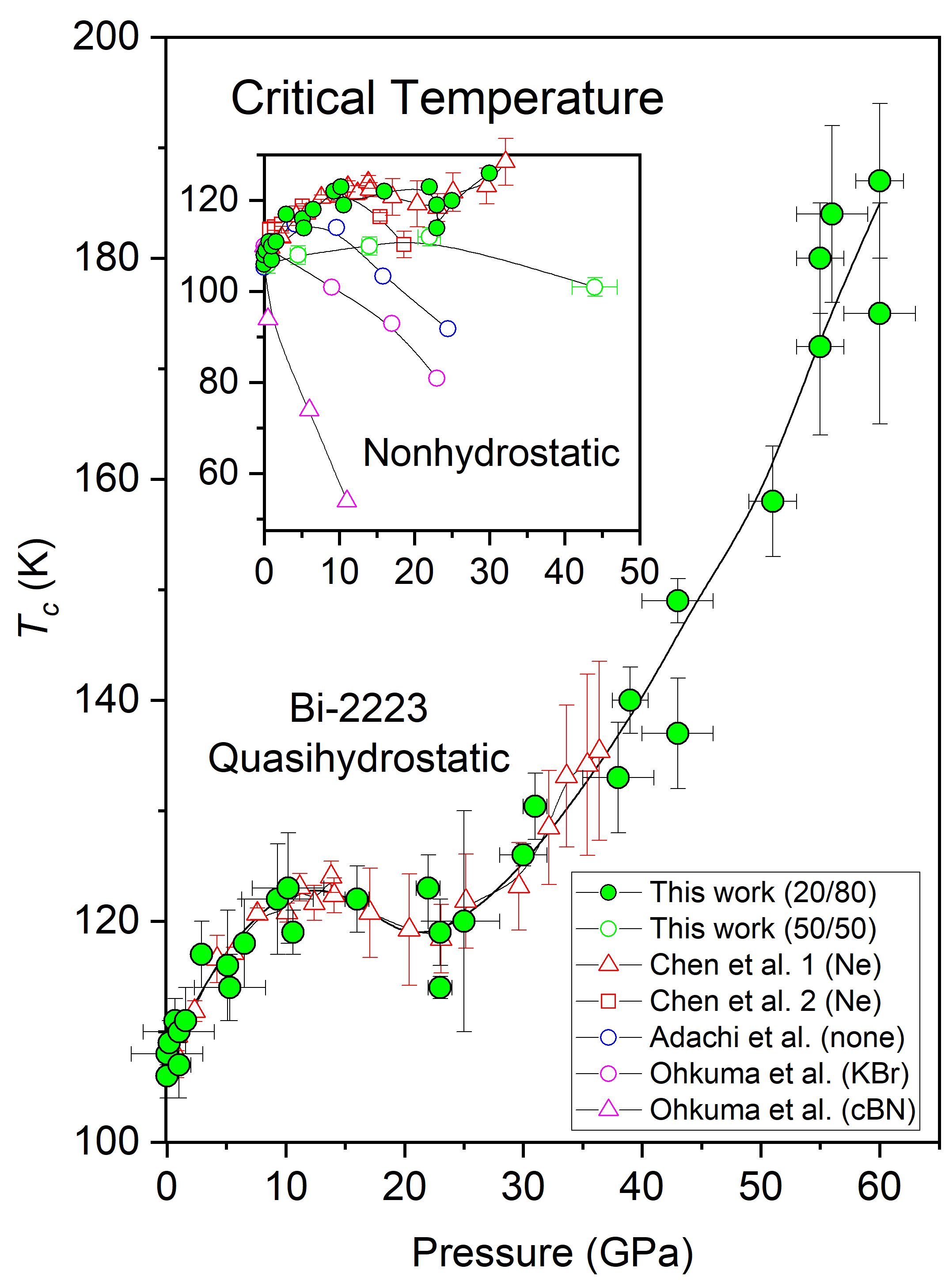}
\caption{Pressure dependence of $T_c$ for Bi-2223 20/80 Bi-2223/KCl mixture, green circles) compared to previous results (open red and blue triangles, circles, and squares) Inset: The results from 50/50 mixture (green, open circles) and previous data in various media labeled nonhydrostatic. experimental data from Chen et al. \cite{chen_enhancement_2010}, Adachi et al. \cite{adachi_pressure_2019} and Ohkuma et al. \cite{ohkuma_probing_2026}. Lines are guides to the eye.}
\label{Fig:P-Tc} 
\end{figure}

Figure \ref{Fig:P-Tc} shows  the pressure dependence of $T_c$ for the three sets of measurements on Bi-2223, including samples compressed with varying degrees of quasihydrostaticity and previous measurements in various compression environments. Below 20 GPa all samples exhibit variations of a dome-like $T_c$ pressure dependence. Our results for quasihydrostatically compressed Bi-2223 agree with data from Chen et al. \cite{chen_enhancement_2010}, whereas less quasihydrostatic (50/50) compression results in skewed $T_c$-$P$ domes with lower maxima and no increase in $T_c$ at the highest pressures, similar to Refs. \cite{adachi_pressure_2019, ohkuma_probing_2026}.

The variability in the pressure dependence of $T_c$ when compressed in different media is consistent with the observation that the structure and electronic properties of the Bi-bearing cuprates are highly sensitive to anisotropic stress \cite{mark_structure_2023}. We therefore attribute differences found for the different Bi-2223/KCl mixtures and with that of others \cite{adachi_pressure_2019, ohkuma_probing_2026} to the compression environment, of these samples, e.g., stiff media such as cubic boron nitride shearing the samples and even destroying superconductivity at confining pressures as low as 10 GPa \cite{ohkuma_probing_2026, xing_effect_2026}. 

Our results have implications for the long-standing paradigm of hole doped cuprate superconductors  \cite{chu_hole-doped_2015}. A notable observation of results obtained from a broad range of cuprates is the parabolic $T_c$ - $n_H$ relation \cite{presland_general_1991}, given as,

\begin{equation}
\frac{T_c}{T_{c,max}}=1-82.6(n_H - n_{H,op})^2,
\label{eq:Tc-nH}
\end{equation}

\noindent with a maximum system dependent $T_{c,max}$ at an optimal $n_{H,op} = 0.16$. On the other hand, the ubiquity of Eq. \ref{eq:Tc-nH} has been challenged with the discoveries of superconductivity in $\mathrm{Sr_2CuO_{4-\delta}}$ with a $T_{c,max}$ = 95 K at $n_H \approx 0.8$ \cite{liu_enhancement_2006, hiroi_updated_2013} and $\mathrm{Ba_2CuO_{3+\delta}}$ with a $T_{c,max}$ = 70 K at $n_H \approx 0.9$ \cite{jin_two-band_2021}. Growing similarly overdoped bulk cuprates by adjusting O doping ($\delta$) is technically difficult \cite{lin_high-quality_2010} and as a result there are limited studies on cuprates with $n_H$ $>$ 0.3. 

Experiments suggest the primary effect of pressure on the electronic structure of cuprates to be an increase in $n_H$ at a constant $\delta$, decoupling the carrier concentration from oxygen doping. Applied pressure, therefore, has been proposed as a method for accessing the highly overdoped region of the phase diagram \cite{jorgensen_pressure-induced_1990, kendziora_polarized_1999, neumeier_pressure_1993, almasan_pressure_1992, ambrosch-draxl_pressure-induced_2004, sakakibara_first-principles_2013}. This pressure induced doping is achieved by transferring charge from the $\mathrm{CuO_2}$ planes to the surrounding charge reservoir layers upon compression \cite{neumeier_hole_1989}. Indeed, the pressure dependence of $T_c$ closely tracks Eq. \ref{eq:Tc-nH} for nearly all cuprates, with select samples exhibiting an enhancement of $T_c$ at higher pressures \cite{mark_progress_2022}.

\begin{figure}
\includegraphics[width=0.5\textwidth]{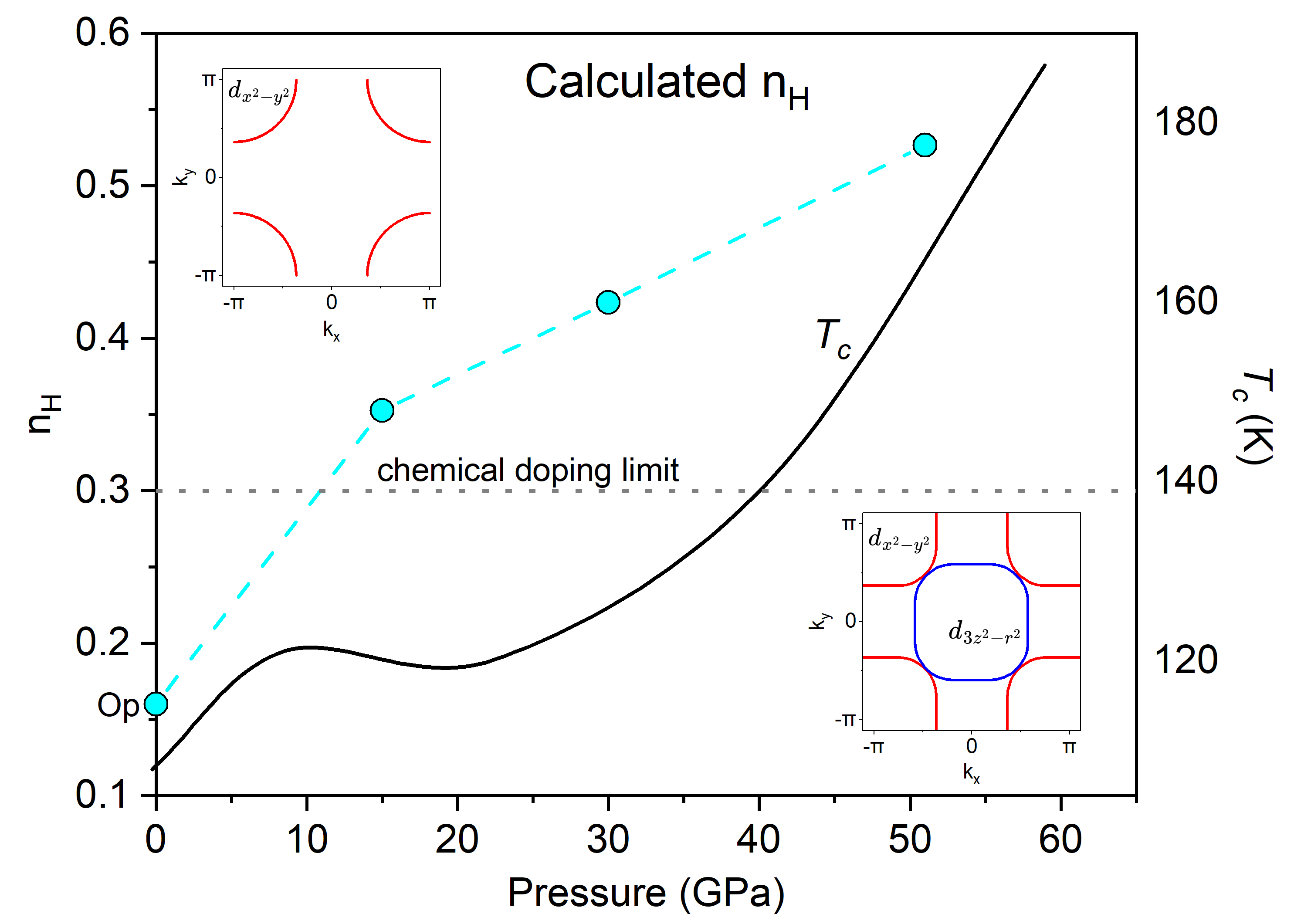}
\caption{Calculated equivalent hole doping (oxygen and pressure-induced) per $\mathrm{CuO_2}$ plane of $\mathrm{Bi_2Sr_2Ca_2Cu_3O_{10+\delta}}$ ($\delta = 0.16$) as a function of pressure from DFT-PBE calculations using experimental unit cell parameters from Ref. \cite{mark_structure_2023} (cyan). The ambient pressure optimal doping limit (Op), and the maximum $n_H$ achievable through chemical doping at ambient pressure are indicated.  Schematic Fermi surfaces for Bi-2223 at comparable hole doping levels adapted from Refs. \cite{maier_two_2019, maier_overdoped_2020} show the ambient pressure Fermi surface composed of the single $d_{x^2-y^2}$ band, and the high-pressure Fermi surface composed of the $d_{x^2-y^2}$ and $d_{3z^2-r^2}$ bands. $T_c$ trace from Figure \ref{Fig:P-Tc} showing enhanced $T_c$ values in the high-doping regime.}
\label{Fig:DFT} 
\end{figure}

In order to better understand the extent of pressure induced hole doping into the $\mathrm{CuO_2}$ planes, DFT-PBE calculations were carried out using the experimentally derived lattice parameters from Ref. \cite{mark_structure_2023}. While DFT-PBE is inadequate for quantitative analysis of the cuprates' electronic structure near the Fermi level \cite{pokharel_sensitivity_2022}, our aim in these calculations is to obtain a qualitative trend with pressure and leave more precise calculations for future work. Figure \ref{Fig:DFT} shows the calculated hole concentration at the Fermi surface per $\mathrm{CuO_2}$ unit at various pressures for optimally doped Bi-2223. As pressure is decoupling $n_H$ from $\delta$ by transferring charge from the $\mathrm{CuO_2}$ planes to the surrounding charge reservoir, significantly higher $n_H$ levels can exist in cuprates under pressure. These calculations suggest an effective doping of $n_H \approx 0.52$, resulting in highly overdoped samples that may exhibit features of the proposed second superconducting dome that has been predicted to correspond with a Lifshitz transition as the planar copper varies from $\mathrm{Cu^{+2}}$ to $\mathrm{Cu^{+3}}$ with charge transfer, resulting in the $d_{3z^2-r^2}$ band crossing Fermi level \cite{maier_two_2019}. We note that a gradual change in compression mechanism in Bi-2223 coincides with this inflection point in the pressure dependence of $T_c$ \cite{mark_structure_2023}.

The deviation from the parabolic trend of Eq. \ref{eq:Tc-nH} under hydrostatic pressure suggests that additional mechanisms influence superconductivity in these materials at high $n_H$. Chen et al. \cite{chen_enhancement_2010} originally proposed this high-pressure regime was due to inner-plane $n_H$ optimization in Bi-2223. Inner plane doping optimization likely plays a role due to pressure reducing the charge imbalance between the inequivalent inner and outer planes \cite{ideta_enhanced_2010}. The observation of non-monotonic behavior in cuprates that lack an inequivalent inner $\mathrm{CuO_2}$ plane, e.g., $\mathrm{Bi_2Sr_2CuO_{6+\delta}}$ (Bi-2201) and $\mathrm{Bi_2Sr_2CaCu_2O_{8+\delta}}$ (Bi-2212)  \cite{deng_higher_2019}, suggests that it is not the only cause of enhanced $T_c$ under pressure.

On the basis of experiments on Bi-2201 and Bi-2212, Deng et al. \cite{deng_higher_2019} suggest pressure doping induces a Lifshitz transition which is responsible for the enhanced $T_c$. Corroborating Raman measurements on all three classes of compounds (Bi-2201, Bi-2212, Bi-2223) suggest that charge redistribution occurs as a function of pressure \cite{chen_enhancement_2010, deng_higher_2019}, consistent with this hypothesis. Analogously, similar non-monotonic $T_c$ pressure dependencies have been reported in the oxides $\mathrm{Ti_4Co_2O}$ \cite{shi_two_nodate} and $\mathrm{MgTi_2O_4}$ \cite{shi_pressure-induced_2025} which have been attributed to electronic restructuring under pressure.

Our results suggest that pressure induced doping drives a Lifshitz transition and results in a second superconducting - $n_H$ dome at high hole dopings, similar to that proposed in $\mathrm{Sr_2CuO_{4-\delta}}$  \cite{geballe_enhanced_2009}. Calculations suggest that at high doping levels the Bi-2223 Fermi surface undergoes a reorganization to a state that allows for significantly more pairing strength in both the d-wave and $\mathrm{s^{\pm}}$ channels and, subsequently, higher $T_c$ \cite{maier_two_2019, oles_orbital_2019}. A rearrangement of the cuprate electronic structure at a sufficiently high $n_H$ is supported by the observation of discontinuities in the strong phonon renormalization effects apparent in Raman spectra of Bi-2223 under pressure \cite{chen_enhancement_2010}, quantities known to couple to the electron density of states \cite{zhou_plane_1997, hadjiev_strong_1998}. Finally, a nodeless superconducting gap in highly overdoped ($n_H > 0.3$) Bi-2212 monolayers has been observed using transmission electron microscopy (TEM); these results demonstrate a change in Fermi surface topology on samples with large strain induced by the substrate, and an accompanying enhanced $T_c$ that may parallel our high-pressure results \cite{zhong_nodeless_2016, jiang_nodeless_2018}.


\section{Conclusions}
\label{Conclusions}
\vspace{\baselineskip}

Extending high-pressure measurements of superconductivity of Bi-2223 to 60 GPa, well above the anomalous non-monotonic $T_c$ regime, the critical temperature rises steeply to 190 K at the highest pressures. The measurements were conducted using three different probes of the Meissner effect. Our data are consistent with proposals of the primary effect of quasihydrostatic pressure being to increase $n_H$ at a constant stoichiometry to access multiple electronic states in unconventional superconductors \cite{das_two_2016, maier_overdoped_2020}. The results suggest that a restructuring of the Fermi surface as a function of $n_H$ leads to a second superconducting regime with a significantly higher $T_c$ well above that observed at ambient pressure. Future work should include further magnetic flux trapping measurements, experiments to still higher pressures, and studies of the possible recoverability and retention of high critical temperatures to lower, including ambient, pressure \cite{deng_ambient-pressure_2026}.

\begin{acknowledgements}
We thank L. Deng, C. W. Chu, X-J. Chen, and F. Restrepo for helpful discussions, and R. Dojutrek and D. Kuntzleman for assistance designing and fabricating much of the equipment used in these experiments. This work was supported by the U.S. National Science Foundation [grant DMR-2104881] and DOE-NNSA [cooperative agreement DE-NA0004153] (Chicago/DOE Alliance Center). N.K. Man acknowledges financial support from the Hanoi University of Science and Technology (HUST) Project Code: T2018-PC-221. Work in the Materials Science Division of Argonne National Laboratory was
supported by the U.S. Department of Energy, Office of Science, Basic Energy Sciences, Materials
Sciences and Engineering Division. This work is a contribution of the George W. Crabtree Institute of Discovery at UIC and ANL.
\end{acknowledgements}

\begin{Data availability}
\small{\noindent \textbf{Data Availability} All data supporting this study has been uploaded to an \href{https://zenodo.org/records/21072112}{online repository} \cite{mark_raw_2026}}. 
\end{Data availability}

\FloatBarrier
\begingroup
\raggedright
\printbibliography

@article{chu_hole-doped_2015,
	series = {Superconducting {Materials}: {Conventional}, {Unconventional} and {Undetermined}},
	title = {Hole-doped cuprate high temperature superconductors},
	volume = {514},
	issn = {0921-4534},
	url = {http://www.sciencedirect.com/science/article/pii/S0921453415000878},
	doi = {10.1016/j.physc.2015.02.047},
	language = {en},
	urldate = {2020-05-28},
	journal = {Physica C},
	author = {Chu, C. W. and Deng, L. Z. and Lv, B.},
	month = jul,
	year = {2015},
	pages = {290--313},
}

@article{timofeev_growth_1998,
	title = {Growth defects in {BSCCO} (2212) single crystal whiskers},
	volume = {309},
	issn = {0921-4534},
	url = {https://www.sciencedirect.com/science/article/pii/S0921453498005620},
	doi = {10.1016/S0921-4534(98)00562-0},
	language = {en},
	number = {1},
	urldate = {2022-11-29},
	journal = {Physica C},
	author = {Timofeev, V. N. and Gorlova, I. G.},
	month = dec,
	year = {1998},
	pages = {113--119},
}

@article{zhou_quantum_2022,
	title = {Quantum phase transition from superconducting to insulating-like state in a pressurized cuprate superconductor},
	volume = {18},
	copyright = {2022 The Author(s)},
	issn = {1745-2481},
	url = {https://www.nature.com/articles/s41567-022-01513-2},
	doi = {10.1038/s41567-022-01513-2},
	language = {en},
	urldate = {2022-03-04},
	journal = {Nat. Phys.},
	author = {Zhou, Yazhou and Guo, Jing and Cai, Shu and Zhao, Jinyu and Gu, Genda and Lin, Chengtian and Yan, Hongtao and Huang, Cheng and Yang, Chongli and Long, Sijin and Gong, Yu and Li, Yanchun and Li, Xiaodong and Wu, Qi and Hu, Jiangping and Zhou, Xingjiang and Xiang, Tao and Sun, Liling},
	month = feb,
	year = {2022},
	pages = {406--410},
}

@article{almasan_pressure_1992,
	title = {Pressure dependence of {T}$_{\textrm{c}}$ and charge transfer in {YBa}$_{\textrm{2}}${Cu}$_{\textrm{3}}${O}$_{\textrm{x}}$(6.35 {\textless} x {\textless} 7) single crystals},
	volume = {69},
	url = {https://link.aps.org/doi/10.1103/PhysRevLett.69.680},
	doi = {10.1103/PhysRevLett.69.680},
	number = {4},
	urldate = {2020-08-24},
	journal = {Phys. Rev. Lett.},
	author = {Almasan, C. C. and Han, S. H. and Lee, B. W. and Paulius, L. M. and Maple, M. B. and Veal, B. W. and Downey, J. W. and Paulikas, A. P. and Fisk, Z. and Schirber, J. E.},
	month = jul,
	year = {1992},
	pages = {680--683},
}

@article{deng_higher_2019,
	title = {Higher superconducting transition temperature by breaking the universal pressure relation},
	volume = {116},
	issn = {0027-8424},
	url = {https://www.ncbi.nlm.nih.gov/pmc/articles/PMC6369809/},
	doi = {10.1073/pnas.1819512116},
	number = {6},
	urldate = {2020-04-20},
	journal = {Proc. Natl. Acad. Sci.},
	author = {Deng, Liangzi and Zheng, Yongping and Wu, Zheng and Huyan, Shuyuan and Wu, Hung-Cheng and Nie, Yifan and Cho, Kyeongjae and Chu, Ching-Wu},
	month = feb,
	year = {2019},
	pages = {2004--2008},
}

@article{jorgensen_pressure-induced_1990,
	title = {Pressure-induced charge transfer and {dT}$_{\textrm{c}}$/{dP} in {YBa}$_{\textrm{2}}${Cu}$_{\textrm{3}}${O}$_{\textrm{7−x}}$},
	volume = {171},
	issn = {0921-4534},
	url = {http://www.sciencedirect.com/science/article/pii/092145349090460V},
	doi = {10.1016/0921-4534(90)90460-V},
	language = {en},
	number = {1},
	urldate = {2020-08-24},
	journal = {Physica C},
	author = {Jorgensen, J. D. and Pei, Shiyou and Lightfoot, P. and Hinks, D. G. and Veal, B. W. and Dabrowski, B. and Paulikas, A. P. and Kleb, R. and Brown, I. D.},
	month = oct,
	year = {1990},
	pages = {93--102},
}

@article{sakakibara_first-principles_2013,
	title = {First-principles band structure and {FLEX} approach to the pressure effect on {T}$_{\textrm{c}}$ of the cuprate superconductors},
	volume = {454},
	issn = {1742-6596},
	url = {http://arxiv.org/abs/1211.1805},
	doi = {10.1088/1742-6596/454/1/012021},
	urldate = {2020-05-25},
	journal = {J. Phys. Conf. Ser.},
	author = {Sakakibara, Hirofumi and Suzuki, Katsuhiro and Usui, Hidetomo and Kuroki, Kazuhiko and Arita, Ryotaro and Scalapino, Douglas J. and Aoki, Hideo},
	month = aug,
	year = {2013},
	pages = {012021},
}

@article{kendziora_polarized_1999,
	title = {Polarized {Electronic} {Raman} {Scattering} {Studies} of {Underdoped} and {Overdoped} {High}-{T}$_{\textrm{c}}$ {Superconductors}},
	volume = {117},
	url = {https://link.springer.com/article/10.1023/A:1022505819394},
	doi = {10.1023/a:1022505819394},
	number = {5-6},
	journal = {J. Low Temp. Phys.},
	author = {Kendziora, C. and Pelloquin, D. and Villard, G.},
	year = {1999},
	pages = {1007--1011},
}

@article{adachi_pressure_2019,
	title = {Pressure effect in {Bi}-2212 and {Bi}-2223 cuprate superconductor},
	volume = {12},
	issn = {1882-0786},
	url = {https://doi.org/10.7567%2F1882-0786%2Fab0521},
	doi = {10.7567/1882-0786/ab0521},
	language = {en},
	number = {4},
	urldate = {2020-07-28},
	journal = {Appl. Phys. Express},
	publisher = {IOP Publishing},
	author = {Adachi, Shintaro and Matsumoto, Ryo and Hara, Hiroshi and Saito, Yoshito and Song, Peng and Takeya, Hiroyuki and Watanabe, Takao and Takano, Yoshihiko},
	month = mar,
	year = {2019},
	pages = {043002},
}

@article{bednorz_possible_1986,
	title = {Possible high {T}$_{\textrm{c}}$ superconductivity in the {Ba}-{La}-{Cu}-{O} system},
	volume = {64},
	issn = {1431-584X},
	url = {https://doi.org/10.1007/BF01303701},
	doi = {10.1007/BF01303701},
	language = {en},
	number = {2},
	urldate = {2019-12-24},
	journal = {Z. Phys. B},
	author = {Bednorz, J. G. and Müller, K. A.},
	month = jun,
	year = {1986},
	pages = {189--193},
}

@article{chu_superconductivity_1987,
	chapter = {Reports},
	title = {Superconductivity at 52.5 {K} in the {Lanthanum}-{Barium}-{Copper}-{Oxide} {System}},
	volume = {235},
	copyright = {1987 by the American Association for the Advancement of Science.},
	issn = {0036-8075, 1095-9203},
	url = {https://science.sciencemag.org/content/235/4788/567},
	doi = {10.1126/science.235.4788.567},
	language = {en},
	number = {4788},
	urldate = {2020-05-28},
	journal = {Science},
	publisher = {American Association for the Advancement of Science},
	author = {Chu, C. W. and Hor, P. H. and Meng, R. L. and Gao, L. and Huang, Z. J.},
	month = jan,
	year = {1987},
	pages = {567--569},
}

@article{presland_general_1991,
	title = {General trends in oxygen stoichiometry effects on {T}$_{\textrm{c}}$ in {Bi} and {Tl} superconductors},
	volume = {176},
	issn = {0921-4534},
	url = {http://www.sciencedirect.com/science/article/pii/0921453491907009},
	doi = {10.1016/0921-4534(91)90700-9},
	language = {en},
	number = {1},
	urldate = {2020-05-05},
	journal = {Physica C},
	author = {Presland, M. R. and Tallon, J. L. and Buckley, R. G. and Liu, R. S. and Flower, N. E.},
	month = may,
	year = {1991},
	pages = {95--105},
}

@article{chen_enhancement_2010,
	title = {Enhancement of superconductivity by pressure-driven competition in electronic order},
	volume = {466},
	url = {https://www.nature.com/articles/nature09293},
	doi = {https://doi.org/10.1038/nature09293},
	urldate = {2020-01-23},
	journal = {Nature},
	author = {Chen, Xiao-Jia and Struzhkin, Viktor V. and Yu, Young and Goncharov, Alexander F. and Lin, Cheng-Tian and Mao, Ho-kwang and Hemley, R. J.},
	year = {2010},
	pages = {950--953},
}

@article{neumeier_hole_1989,
	title = {Hole filling and pair breaking by {Pr} ions in {YBa}$_{\textrm{2}}${Cu}$_{\textrm{3}}${O}$_{\textrm{6.95±0.02}}$},
	volume = {63},
	url = {https://link.aps.org/doi/10.1103/PhysRevLett.63.2516},
	doi = {10.1103/PhysRevLett.63.2516},
	number = {22},
	urldate = {2022-02-01},
	journal = {Phys. Rev. Lett.},
	publisher = {American Physical Society},
	author = {Neumeier, J. J. and Bjørnholm, T. and Maple, M. B. and Schuller, Ivan K.},
	month = nov,
	year = {1989},
	pages = {2516--2519},
}

@article{hiroi_updated_2013,
	title = {“{Updated}” {World} {Record} for {Superconducting} {Transition} {Temperature}},
	volume = {10},
	issn = {2188-076X},
	url = {https://journals.jps.jp/doi/abs/10.7566/JPSJNC.10.04},
	doi = {10.7566/JPSJNC.10.04},
	urldate = {2020-11-03},
	journal = {JPSJ News Comments},
	publisher = {The Physical Society of Japan},
	author = {Hiroi, Zenji},
	month = jan,
	year = {2013},
	pages = {04},
}

@article{klotz_hydrostatic_1993,
	title = {Hydrostatic pressure dependence of the superconducting transition temperature to 7 {GPa} in {Bi}$_{\textrm{2}}${Ca}$_{\textrm{1}}${Sr}$_{\textrm{2}}${Cu}$_{\textrm{2}}${O}$_{\textrm{8+y}}$ as a function of oxygen content},
	volume = {209},
	issn = {0921-4534},
	url = {https://www.sciencedirect.com/science/article/pii/0921453493905669},
	doi = {10.1016/0921-4534(93)90566-9},
	language = {en},
	number = {4},
	urldate = {2021-12-03},
	journal = {Physica C},
	author = {Klotz, S. and Schilling, J. S.},
	month = may,
	year = {1993},
	pages = {499--506},
}

@article{perdew_generalized_1996,
	title = {Generalized {Gradient} {Approximation} {Made} {Simple}},
	volume = {77},
	issn = {1079-7114},
	url = {https://journals.aps.org/prl/abstract/10.1103/PhysRevLett.77.3865},
	doi = {10.1103/PhysRevLett.77.3865},
	language = {eng},
	number = {18},
	journal = {Phys. Rev. Lett.},
	author = {Perdew, John P. and Burke, Kieron and Ernzerhof, Matthias},
	month = oct,
	year = {1996},
	pages = {3865--3868},
}

@article{gregoryanz_superconductivity_2002,
	title = {Superconductivity in the chalcogens up to multimegabar pressures},
	volume = {65},
	issn = {0163-1829, 1095-3795},
	url = {https://link.aps.org/doi/10.1103/PhysRevB.65.064504},
	doi = {10.1103/PhysRevB.65.064504},
	language = {en},
	number = {6},
	urldate = {2020-06-02},
	journal = {Phys. Rev. B},
	author = {Gregoryanz, Eugene and Struzhkin, Viktor and Hemley, Russell and Eremets, Mikhail and Mao, Ho-kwang and Timofeev, Yuri},
	month = jan,
	year = {2002},
	pages = {064504},
}

@article{ideta_enhanced_2010,
	title = {Enhanced {Superconducting} {Gaps} in the {Trilayer} {High}-{Temperature} {Bi}$_{\textrm{2}}${Sr}$_{\textrm{2}}${Ca}$_{\textrm{2}}${Cu}$_{\textrm{3}}${O}$_{\textrm{10+δ}}$ {Cuprate} {Superconductor}},
	volume = {104},
	issn = {0031-9007, 1079-7114},
	url = {https://link.aps.org/doi/10.1103/PhysRevLett.104.227001},
	doi = {10.1103/PhysRevLett.104.227001},
	language = {en},
	number = {22},
	urldate = {2020-06-02},
	journal = {Phys. Rev. Lett.},
	author = {Ideta, S. and Takashima, K. and Hashimoto, M. and Yoshida, T. and Fujimori, A. and Anzai, H. and Fujita, T. and Nakashima, Y. and Ino, A. and Arita, M. and Namatame, H. and Taniguchi, M. and Ono, K. and Kubota, M. and Lu, D. H. and Shen, Z.-X. and Kojima, K. M. and Uchida, S.},
	month = jun,
	year = {2010},
	pages = {227001},
}

@article{ambrosch-draxl_pressure-induced_2004,
	title = {Pressure-induced hole doping of the {Hg}-based cuprate superconductors},
	volume = {92},
	issn = {0031-9007, 1079-7114},
	url = {http://arxiv.org/abs/cond-mat/0408288},
	doi = {10.1103/PhysRevLett.92.187004},
	number = {18},
	urldate = {2020-05-25},
	journal = {Phys. Rev. Lett.},
	author = {Ambrosch-Draxl, C. and Sherman, E. Ya and Auer, H. and Thonhauser, T.},
	month = may,
	year = {2004},
	pages = {187004},
}

@article{chu_evidence_1987,
	title = {Evidence for superconductivity above 40 {K} in the {La}-{Ba}-{Cu}-{O} compound system},
	volume = {58},
	url = {https://link.aps.org/doi/10.1103/PhysRevLett.58.405},
	doi = {10.1103/PhysRevLett.58.405},
	number = {4},
	urldate = {2020-05-28},
	journal = {Phys. Rev. Lett.},
	publisher = {American Physical Society},
	author = {Chu, C. W. and Hor, P. H. and Meng, R. L. and Gao, L. and Huang, Z. J. and Wang, {and} Y. Q.},
	month = jan,
	year = {1987},
	pages = {405--407},
}

@article{neumeier_pressure_1993,
	title = {Pressure dependence of the superconducting transition temperature of {YBa}$_{\textrm{2}}${Cu}$_{\textrm{3}}${O}$_{\textrm{7}}$ as a function of carrier concentration: {A} test for a simple charge-transfer model},
	volume = {47},
	issn = {0163-1829, 1095-3795},
	shorttitle = {Pressure dependence of the superconducting transition temperature of {YBa} 2 {Cu} 3 {O} 7 as a function of carrier concentration},
	url = {https://link.aps.org/doi/10.1103/PhysRevB.47.8385},
	doi = {10.1103/PhysRevB.47.8385},
	language = {en},
	number = {13},
	urldate = {2020-07-22},
	journal = {Phys. Rev. B},
	author = {Neumeier, J. J. and Zimmermann, H. A.},
	month = apr,
	year = {1993},
	pages = {8385--8388},
}

@article{mito_uniaxial_2017,
	title = {Uniaxial strain effects on the superconducting transition in {Re}-doped {Hg}-1223 cuprate superconductors},
	volume = {95},
	url = {https://link.aps.org/doi/10.1103/PhysRevB.95.064503},
	doi = {10.1103/PhysRevB.95.064503},
	number = {6},
	urldate = {2020-07-28},
	journal = {Phys. Rev. B},
	publisher = {American Physical Society},
	author = {Mito, Masaki and Ogata, Kazuma and Goto, Hiroki and Tsuruta, Kazuki and Nakamura, Kazuma and Deguchi, Hiroyuki and Horide, Tomoya and Matsumoto, Kaname and Tajiri, Takayuki and Hara, Hiroshi and Ozaki, Toshinori and Takeya, Hiroyuki and Takano, Yoshihiko},
	month = feb,
	year = {2017},
	pages = {064503},
}

@article{adachi_electrical_2001,
	title = {Electrical resistivity measurements on fragile organic single crystals in the diamond anvil cell},
	volume = {72},
	issn = {0034-6748},
	url = {https://aip.scitation.org/doi/abs/10.1063/1.1361085},
	doi = {10.1063/1.1361085},
	number = {5},
	urldate = {2022-05-22},
	journal = {Rev. Sci. Instrum.},
	publisher = {American Institute of Physics},
	author = {Adachi, T. and Tanaka, H. and Kobayashi, H. and Miyazaki, T.},
	month = may,
	year = {2001},
	pages = {2358--2360},
}

@article{mark_progress_2022,
	title = {Progress and prospects for cuprate high temperature superconductors under pressure},
	volume = {42},
	copyright = {All rights reserved},
	issn = {0895-7959},
	url = {https://doi.org/10.1080/08957959.2022.2059366},
	doi = {10.1080/08957959.2022.2059366},
	number = {2},
	urldate = {2022-08-09},
	journal = {High Press. Res.},
	author = {Mark, Alexander C. and Campuzano, Juan Carlos and Hemley, Russell J.},
	month = apr,
	year = {2022},
	pages = {137--199},
}

@article{timofeev_inductive_1999,
	title = {Inductive method for investigation of ferromagnetic properties of materials under pressure},
	volume = {70},
	issn = {0034-6748},
	url = {https://doi.org/10.1063/1.1150036},
	doi = {10.1063/1.1150036},
	number = {10},
	urldate = {2023-08-21},
	journal = {Rev. Sci. Instrum.},
	author = {Timofeev, Yuri A. and Mao, Ho-kwang and Struzhkin, Viktor V. and Hemley, Russell J.},
	month = oct,
	year = {1999},
	pages = {4059--4061},
}

@article{mark_structure_2023,
	title = {Structure and equation of state of {Bi}$_{\textrm{2}}${Sr}$_{\textrm{2}}${Ca}$_{\textrm{n-1}}${Cu}$_{\textrm{n}}${O}$_{\textrm{2n+4+δ}}$ from x-ray diffraction to megabar pressures},
	volume = {7},
	copyright = {All rights reserved},
	url = {https://link.aps.org/doi/10.1103/PhysRevMaterials.7.064803},
	doi = {10.1103/PhysRevMaterials.7.064803},
	number = {6},
	urldate = {2023-06-14},
	journal = {Phys. Rev. Mater.},
	publisher = {American Physical Society},
	author = {Mark, Alexander C. and Ahart, Muhtar and Kumar, Ravhi and Park, Changyong and Meng, Yue and Popov, Dmitry and Deng, Liangzi and Chu, Ching-Wu and Campuzano, Juan Carlos and Hemley, Russell J.},
	month = jun,
	year = {2023},
	pages = {064803},
}

@article{timofeev_improved_2002,
	title = {Improved techniques for measurement of superconductivity in diamond anvil cells by magnetic susceptibility},
	volume = {73},
	issn = {0034-6748},
	url = {http://aip.scitation.org/doi/10.1063/1.1431257},
	doi = {10.1063/1.1431257},
	number = {2},
	urldate = {2021-03-11},
	journal = {Rev. Sci. Instrum.},
	publisher = {American Institute of Physics},
	author = {Timofeev, Yuri A. and Struzhkin, Viktor V. and Hemley, Russell J. and Mao, Ho-kwang and Gregoryanz, Eugene A.},
	month = feb,
	year = {2002},
	pages = {371--377},
}

@article{deemyad_dependence_2003,
	title = {Dependence of the superconducting transition temperature of single and polycrystalline {MgB}$_{\textrm{2}}$ on hydrostatic pressure},
	volume = {385},
	issn = {0921-4534},
	url = {https://www.sciencedirect.com/science/article/pii/S0921453402023006},
	doi = {10.1016/S0921-4534(02)02300-6},
	number = {1},
	urldate = {2024-03-18},
	journal = {Physica C},
	author = {Deemyad, S. and Tomita, T. and Hamlin, J. J. and Beckett, B. R. and Schilling, J. S. and Hinks, D. G. and Jorgensen, J. D. and Lee, S. and Tajima, S.},
	month = mar,
	year = {2003},
	pages = {105--116},
}

@incollection{goldfarb_alternating-field_1991,
	address = {Boston, MA},
	title = {Alternating-{Field} {Susceptometry} and {Magnetic} {Susceptibility} of {Superconductors}},
	isbn = {978-1-4899-2379-0},
	url = {https://doi.org/10.1007/978-1-4899-2379-0_3},
	doi = {10.1007/978-1-4899-2379-0_3},
	language = {en},
	urldate = {2024-09-09},
	booktitle = {Magnetic {Susceptibility} of {Superconductors} and {Other} {Spin} {Systems}},
	publisher = {Springer US},
	author = {Goldfarb, R. B. and Lelental, M. and Thompson, C. A.},
	editor = {Hein, Robert A. and Francavilla, Thomas L. and Liebenberg, Donald H.},
	year = {1991},
	pages = {49--80},
}

@article{maier_two_2019,
	title = {Two pairing domes as {Cu}$^{\textrm{2+}}$ varies to {Cu}$^{\textrm{3+}}$},
	volume = {99},
	url = {https://link.aps.org/doi/10.1103/PhysRevB.99.224515},
	doi = {10.1103/PhysRevB.99.224515},
	number = {22},
	urldate = {2024-09-24},
	journal = {Phys. Rev. B},
	publisher = {American Physical Society},
	author = {Maier, Thomas and Berlijn, T. and Scalapino, D. J.},
	month = jun,
	year = {2019},
	pages = {224515},
}

@article{geballe_enhanced_2009,
	title = {Enhanced superconductivity in {Sr}$_{\textrm{2}}${CuO}$_{\textrm{4−\textit{v}}}$},
	volume = {469},
	issn = {0921-4534},
	url = {https://www.sciencedirect.com/science/article/pii/S0921453409000598},
	doi = {10.1016/j.physc.2009.03.054},
	number = {13},
	urldate = {2024-09-25},
	journal = {Physica C: Superconductivity},
	author = {Geballe, T. H. and Marezio, M.},
	month = jul,
	year = {2009},
	pages = {680--684},
}

@article{dewaele_compression_2008,
	title = {Compression curves of transition metals in the {Mbar} range: {Experiments} and projector augmented-wave calculations},
	volume = {78},
	shorttitle = {Compression curves of transition metals in the {Mbar} range},
	url = {https://link.aps.org/doi/10.1103/PhysRevB.78.104102},
	doi = {10.1103/PhysRevB.78.104102},
	number = {10},
	urldate = {2024-11-01},
	journal = {Phys. Rev. B},
	publisher = {American Physical Society},
	author = {Dewaele, Agnès and Torrent, Marc and Loubeyre, Paul and Mezouar, Mohamed},
	month = sep,
	year = {2008},
	pages = {104102},
}

@article{liu_enhancement_2006,
	title = {Enhancement of the superconducting critical temperature of {Sr}$_{\textrm{2}}${CuO}$_{\textrm{3+δ}}$ up to 95 {K} by ordering dopant atoms},
	volume = {74},
	url = {https://link.aps.org/doi/10.1103/PhysRevB.74.100506},
	doi = {10.1103/PhysRevB.74.100506},
	number = {10},
	urldate = {2024-11-07},
	journal = {Phys. Rev. B},
	publisher = {American Physical Society},
	author = {Liu, Q. Q. and Yang, H. and Qin, X. M. and Yu, Y. and Yang, L. X. and Li, F. Y. and Yu, R. C. and Jin, C. Q. and Uchida, S.},
	month = sep,
	year = {2006},
	pages = {100506},
}

@article{maier_overdoped_2020,
	title = {Overdoped end of the cuprate phase diagram},
	volume = {2},
	url = {https://link.aps.org/doi/10.1103/PhysRevResearch.2.033132},
	doi = {10.1103/PhysRevResearch.2.033132},
	number = {3},
	urldate = {2024-11-07},
	journal = {Phys. Rev. Res.},
	publisher = {American Physical Society},
	author = {Maier, Thomas A. and Karakuzu, Seher and Scalapino, Douglas J.},
	month = jul,
	year = {2020},
	pages = {033132},
}

@article{lin_high-quality_2010,
	title = {High-{Quality} {Large}-{Sized} {Single} {Crystals} of {Pb}-{Doped} {Bi}$_{\textrm{2}}${Sr}$_{\textrm{2}}${CuO}$_{\textrm{6+δ}}$ {High}-{T}$_{\textrm{c}}$ {Superconductors} {Grown} with {Traveling} {Solvent} {Floating} {Zone} {Method}},
	volume = {27},
	issn = {0256-307X},
	url = {https://dx.doi.org/10.1088/0256-307X/27/8/087401},
	doi = {10.1088/0256-307X/27/8/087401},
	language = {en},
	number = {8},
	urldate = {2024-11-07},
	journal = {Chinese Phys. Lett.},
	author = {Lin, Zhao and Wen-Tao, Zhang and Hai-Yun, Liu and Jian-Qiaov, Meng and Guo-Dong, Liu and Wei, Lu and Xiao-Li, Dong and Xing-Jiang, Zhou},
	month = aug,
	year = {2010},
	pages = {087401},
}

@article{jiang_nodeless_2018,
	title = {Nodeless {High}-{T}$_{\textrm{c}}$ {Superconductivity} in the {Highly} {Overdoped} {CuO}$_{\textrm{2}}$ {Monolayer}},
	volume = {121},
	url = {https://link.aps.org/doi/10.1103/PhysRevLett.121.227002},
	doi = {10.1103/PhysRevLett.121.227002},
	number = {22},
	urldate = {2024-11-11},
	journal = {Phys. Rev. Lett.},
	publisher = {American Physical Society},
	author = {Jiang, Kun and Wu, Xianxin and Hu, Jiangping and Wang, Ziqiang},
	month = nov,
	year = {2018},
	pages = {227002},
}

@article{das_two_2016,
	title = {Two types of superconducting domes in unconventional superconductors},
	volume = {18},
	issn = {1367-2630},
	url = {https://dx.doi.org/10.1088/1367-2630/18/10/103033},
	doi = {10.1088/1367-2630/18/10/103033},
	language = {en},
	number = {10},
	urldate = {2024-11-11},
	journal = {New J. Phys.},
	publisher = {IOP Publishing},
	author = {Das, Tanmoy and Panagopoulos, Christos},
	month = oct,
	year = {2016},
	pages = {103033},
}

@article{zhong_nodeless_2016,
	title = {Nodeless pairing in superconducting copper-oxide monolayer films on {Bi}$_{\textrm{2}}${Sr}$_{\textrm{2}}${CaCu}$_{\textrm{2}}${O}$_{\textrm{8+δ}}$},
	volume = {61},
	issn = {2095-9273},
	url = {https://www.sciencedirect.com/science/article/pii/S2095927316300494},
	doi = {10.1007/s11434-016-1145-4},
	number = {16},
	urldate = {2024-12-18},
	journal = {Sci. Bull.},
	author = {Zhong, Yong and Wang, Yang and Han, Sha and Lv, Yan-Feng and Wang, Wen-Lin and Zhang, Ding and Ding, Hao and Zhang, Yi-Min and Wang, Lili and He, Ke and Zhong, Ruidan and Schneeloch, John A. and Gu, Gen-Da and Song, Can-Li and Ma, Xu-Cun and Xue, Qi-Kun},
	month = aug,
	year = {2016},
	pages = {1239--1247},
}

@article{man_signature_2024,
	title = {Signature of {T}$_{\textrm{c}}$ above 111 {K} in {Li}-doped triple-layered cuprates {Bi}$_{\textrm{1.6}}${Pb}$_{\textrm{0.4}}${Sr}$_{\textrm{2}}${Ca}$_{\textrm{2}}$({Cu}$_{\textrm{x}}${Li}$_{\textrm{1-x}}$)$_{\textrm{3}}${O}$_{\textrm{10+δ}}$},
	volume = {8},
	url = {https://link.aps.org/doi/10.1103/PhysRevMaterials.8.124802},
	doi = {10.1103/PhysRevMaterials.8.124802},
	number = {12},
	urldate = {2024-12-24},
	journal = {Phys. Rev. Mater.},
	publisher = {American Physical Society},
	author = {Man, N. K. and Do, Huu T.},
	month = dec,
	year = {2024},
	pages = {124802},
}

@article{giannozzi_quantum_2009,
	title = {{QUANTUM} {ESPRESSO}: a modular and open-source software project for quantum simulations of materials},
	volume = {21},
	issn = {0953-8984},
	shorttitle = {{QUANTUM} {ESPRESSO}},
	url = {https://dx.doi.org/10.1088/0953-8984/21/39/395502},
	doi = {10.1088/0953-8984/21/39/395502},
	language = {en},
	number = {39},
	urldate = {2024-12-24},
	journal = {J. Phys.: Condens. Matter},
	author = {Giannozzi, Paolo and Baroni, Stefano and Bonini, Nicola and Calandra, Matteo and Car, Roberto and Cavazzoni, Carlo and Ceresoli, Davide and Chiarotti, Guido L. and Cococcioni, Matteo and Dabo, Ismaila and Corso, Andrea Dal and Gironcoli, Stefano de and Fabris, Stefano and Fratesi, Guido and Gebauer, Ralph and Gerstmann, Uwe and Gougoussis, Christos and Kokalj, Anton and Lazzeri, Michele and Martin-Samos, Layla and Marzari, Nicola and Mauri, Francesco and Mazzarello, Riccardo and Paolini, Stefano and Pasquarello, Alfredo and Paulatto, Lorenzo and Sbraccia, Carlo and Scandolo, Sandro and Sclauzero, Gabriele and Seitsonen, Ari P. and Smogunov, Alexander and Umari, Paolo and Wentzcovitch, Renata M.},
	month = sep,
	year = {2009},
	pages = {395502},
}

@article{giannozzi_advanced_2017,
	title = {Advanced capabilities for materials modelling with {Quantum} {ESPRESSO}},
	volume = {29},
	issn = {0953-8984},
	url = {https://dx.doi.org/10.1088/1361-648X/aa8f79},
	doi = {10.1088/1361-648X/aa8f79},
	language = {en},
	number = {46},
	urldate = {2024-12-24},
	journal = {J. Phys. Condens. Matter.},
	publisher = {IOP Publishing},
	author = {Giannozzi, P. and Andreussi, O. and Brumme, T. and Bunau, O. and Nardelli, M. Buongiorno and Calandra, M. and Car, R. and Cavazzoni, C. and Ceresoli, D. and Cococcioni, M. and Colonna, N. and Carnimeo, I. and Corso, A. Dal and Gironcoli, S. de and Delugas, P. and DiStasio, R. A. and Ferretti, A. and Floris, A. and Fratesi, G. and Fugallo, G. and Gebauer, R. and Gerstmann, U. and Giustino, F. and Gorni, T. and Jia, J. and Kawamura, M. and Ko, H.-Y. and Kokalj, A. and Küçükbenli, E. and Lazzeri, M. and Marsili, M. and Marzari, N. and Mauri, F. and Nguyen, N. L. and Nguyen, H.-V. and Otero-de-la-Roza, A. and Paulatto, L. and Poncé, S. and Rocca, D. and Sabatini, R. and Santra, B. and Schlipf, M. and Seitsonen, A. P. and Smogunov, A. and Timrov, I. and Thonhauser, T. and Umari, P. and Vast, N. and Wu, X. and Baroni, S.},
	month = oct,
	year = {2017},
	pages = {465901},
}

@article{pokharel_sensitivity_2022,
	title = {Sensitivity of the electronic and magnetic structures of cuprate superconductors to density functional approximations},
	volume = {8},
	copyright = {2022 The Author(s)},
	issn = {2057-3960},
	url = {https://www.nature.com/articles/s41524-022-00711-z},
	doi = {10.1038/s41524-022-00711-z},
	language = {en},
	number = {1},
	urldate = {2024-12-24},
	journal = {npj Comput. Mater.},
	publisher = {Nature Publishing Group},
	author = {Pokharel, Kanun and Lane, Christopher and Furness, James W. and Zhang, Ruiqi and Ning, Jinliang and Barbiellini, Bernardo and Markiewicz, Robert S. and Zhang, Yubo and Bansil, Arun and Sun, Jianwei},
	month = feb,
	year = {2022},
	pages = {1--11},
}

@article{prandini_precision_2018,
	title = {Precision and efficiency in solid-state pseudopotential calculations},
	volume = {4},
	copyright = {2018 The Author(s)},
	issn = {2057-3960},
	url = {https://www.nature.com/articles/s41524-018-0127-2},
	doi = {10.1038/s41524-018-0127-2},
	language = {en},
	number = {1},
	urldate = {2024-12-24},
	journal = {npj Comput. Mater.},
	publisher = {Nature Publishing Group},
	author = {Prandini, Gianluca and Marrazzo, Antimo and Castelli, Ivano E. and Mounet, Nicolas and Marzari, Nicola},
	month = dec,
	year = {2018},
	pages = {1--13},
}

@article{struzhkin_superconductivity_1997,
	title = {Superconductivity at 10–17 {K} in compressed sulphur},
	volume = {390},
	copyright = {1997 Macmillan Magazines Ltd.},
	issn = {1476-4687},
	url = {https://www.nature.com/articles/37074},
	doi = {10.1038/37074},
	language = {en},
	number = {6658},
	urldate = {2025-02-11},
	journal = {Nature},
	publisher = {Nature Publishing Group},
	author = {Struzhkin, Viktor V. and Hemley, Russell J. and Mao, Ho-kwang and Timofeev, Yuri A.},
	month = nov,
	year = {1997},
	pages = {382--384},
}

@article{akahama_pressure_2006,
	title = {Pressure calibration of diamond anvil {Raman} gauge to {310GPa}},
	volume = {100},
	issn = {0021-8979},
	url = {https://doi.org/10.1063/1.2335683},
	doi = {10.1063/1.2335683},
	number = {4},
	urldate = {2025-02-12},
	journal = {J. Appl. Phys},
	author = {Akahama, Yuichi and Kawamura, Haruki},
	month = aug,
	year = {2006},
	pages = {043516},
}

@article{man_improvement_2019,
	title = {Improvement of {Critical} {Current} {Density} in {Bi}-2223 {Superconductor} by {Ag}-{Doping}},
	volume = {35},
	copyright = {Copyright (c) 2019 VNU Journal of Science: Mathematics - Physics},
	issn = {2588-1124},
	url = {https://js.vnu.edu.vn/MaP/article/view/4357},
	doi = {10.25073/2588-1124/vnumap.4357},
	language = {en},
	number = {4},
	urldate = {2025-04-02},
	journal = {VNU J. Sci. Math. Phys.},
	author = {Man, Nguyen Khac and Hoa, Nguyen Duc and Nhan, Duong Thi Thanh},
	month = dec,
	year = {2019},
	note = {Number: 4},
}

@article{takano_high-tc_1988,
	title = {High-{T}$_{\textrm{c}}$ {Phase} {Promoted} and {Stabilized} in the {Bi}, {Pb}-{Sr}-{Ca}-{Cu}-{O} {System}},
	volume = {27},
	issn = {1347-4065},
	url = {https://iopscience.iop.org/article/10.1143/JJAP.27.L1041/meta},
	doi = {10.1143/JJAP.27.L1041},
	language = {en},
	number = {6A},
	urldate = {2025-05-27},
	journal = {Jpn. J. Appl. Phys.},
	publisher = {IOP Publishing},
	author = {Takano, Mikio and Takada, Jun and Oda, Kiichi and Kitaguchi, Hitoshi and Miura, Yoshinari and Ikeda, Yasunori and Tomii, Yoichi and Mazaki, Hiromasa},
	month = jun,
	year = {1988},
	pages = {L1041},
}

@article{jin_two-band_2021,
	title = {Two-band conduction and nesting instabilities in superconducting {Ba}$_{\textrm{2}}${CuO}$_{\textrm{3+δ}}$: {First}-principles study},
	volume = {104},
	shorttitle = {Two-band conduction and nesting instabilities in superconducting \$\{{\textbackslash}mathrm\{{Ba}\}\}\_\{2\}\{{\textbackslash}mathrm\{{CuO}\}\}\_\{3+{\textbackslash}ensuremath\{{\textbackslash}delta\}\}\$},
	url = {https://link.aps.org/doi/10.1103/PhysRevB.104.054516},
	doi = {10.1103/PhysRevB.104.054516},
	number = {5},
	urldate = {2025-06-26},
	journal = {Phys. Rev. B},
	publisher = {American Physical Society},
	author = {Jin, Hyo-Sun and Pickett, Warren E. and Lee, Kwan-Woo},
	month = aug,
	year = {2021},
	pages = {054516},
}

@article{dou_effect_1995,
	title = {Effect of silver on phase formation and superconducting properties of {Bi}-2223/{Ag} tapes},
	volume = {5},
	issn = {1558-2515},
	url = {https://ieeexplore.ieee.org/document/402936},
	doi = {10.1109/77.402936},
	number = {2},
	urldate = {2025-07-18},
	journal = {IEEE Trans. Appl. Supercond.},
	author = {Dou, S.X. and Guo, Y.C. and Wang, R.K. and Ionescu, M. and Liu, H.K. and Babic, E. and Kusevic, I.},
	month = jun,
	year = {1995},
	pages = {1830--1833},
}

@article{grivel_effects_1993,
	title = {Effects of {Pb} in the first stages of the {Bi}(2223) phase formation},
	volume = {6},
	issn = {0953-2048},
	url = {https://doi.org/10.1088/0953-2048/6/10/004},
	doi = {10.1088/0953-2048/6/10/004},
	language = {en},
	number = {10},
	urldate = {2026-04-30},
	journal = {Supercond. Sci. Technol.},
	author = {Grivel, J.-C. and Jeremie, A. and Hensel, B. and Flükiger, R.},
	month = oct,
	year = {1993},
	pages = {725},
}

@article{macmanus-driscoll_study_1997,
	title = {Study of ({Bi},{Pb})-2223 phase formation in reaction couples},
	volume = {10},
	issn = {0953-2048},
	url = {https://doi.org/10.1088/0953-2048/10/12/023},
	doi = {10.1088/0953-2048/10/12/023},
	language = {en},
	number = {12},
	urldate = {2026-04-30},
	journal = {Supercond. Sci. Technol.},
	author = {MacManus-Driscoll, J. L. and Yi, Z.},
	month = dec,
	year = {1997},
	pages = {970},
}

@article{gonzalez_arevalo_enhanced_2026,
	title = {Enhanced stability and bulk superconducting properties of {Ag} intercalated {Bi}$_{\textrm{1.6}}${Pb}$_{\textrm{0.4}}${Sr}$_{\textrm{2}}${Ca}$_{\textrm{2}}${Cu}$_{\textrm{3}}${O}$_{\textrm{10+δ}}$},
	volume = {39},
	issn = {0953-2048},
	url = {https://doi.org/10.1088/1361-6668/ae5690},
	doi = {10.1088/1361-6668/ae5690},
	language = {en},
	number = {4},
	urldate = {2026-04-30},
	journal = {Supercond. Sci. Technol.},
	publisher = {IOP Publishing},
	author = {Gonzalez Arevalo, D and Do, Huu T and Mark, Alexander C and Rodriguez, David and Zangeneh, Danial and Klie, Robert F and Hemley, Russell J and Man, N K},
	month = apr,
	year = {2026},
	pages = {045007},
}
\endgroup

\end{document}